\documentclass[aps,prl,twocolumn,showpacs]{revtex4}
\usepackage{graphicx}
\usepackage{longtable}
\usepackage{amssymb,amsmath}

\begin{document}

\title{Early onset of ground-state deformation in the neutron-deficient polonium isotopes}

\author{T.E.~Cocolios$^{1,2}$, W.~Dexters$^1$, M.D.~Seliverstov$^{1,3,4}$, A.N.~Andreyev$^{1,5}$, S.~Antalic$^6$, A.E.~Barzakh$^3$, B.~Bastin$^1$\footnote{Present address: GANIL, France}, J.~B\"uscher$^1$, I.G.~Darby$^1$, D.V.~Fedorov$^3$, V.N.~Fedosseyev$^7$, K.T.~Flanagan$^{8,9}$, S.~Franchoo$^{10}$, S.~Fritzsche$^{11,12}$, G.~Huber$^4$, M.~Huyse$^1$, M.~Keupers$^1$, U.~K{\"o}ster$^{13}$, Yu.~Kudryavtsev$^1$, E.~Man{\'e}$^8$\footnote{Present address: TRIUMF, Canada}, B.A.~Marsh$^7$, P.L.~Molkanov$^3$, R.D.~Page$^{14}$, A.M.~Sjoedin$^{7,15}$, I.~Stefan$^{10}$, J.~Van de Walle$^{1,2}$\footnote{Present address: KVI, The Netherlands}, P.~Van Duppen$^1$, M.~Venhart$^{1,16}$, S.G.~Zemlyanoy$^{17}$, and M.~Bender$^{18}$, P.-H.~Heenen$^{19}$}
\affiliation{$^1$ Instituut voor Kern- en Stralingsfysica, K.U. Leuven, B-3001 Leuven, Belgium}
\affiliation{$^2$ ISOLDE, CERN, CH-1211 Geneva 23, Switzerland}
\affiliation{$^3$ Petersburg Nuclear Physics Institute, 188350 Gatchina, Russia}
\affiliation{$^4$ Institut f\"ur Physik, Johannes Gutenberg Universit{\"a}t, D-55099 Mainz, Germany}
\affiliation{$^5$ School of Engineering and Science, University of West Scotland, Paisley, PA1 2BE, United Kingdom and Scottish Universities Physics Alliance (SUPA)}
\affiliation{$^6$ Department of Physics and Biophysics, Comenius University, Bratislava 842 48, Slovakia}
\affiliation{$^7$ EN Department, CERN, CH-1211 Geneva 23, Switzerland}
\affiliation{$^8$ Department of Physics, University of Manchester, Manchester, M60 1AD, United Kingdom}
\affiliation{$^9$ Centre de Spectrom\'etrie Nucl\'eaire et de Spectrom\'etrie de Masse, F-91405 Orsay, France}
\affiliation{$^{10}$ Institut de Physique Nucl\'eaire d'Orsay, F-91406 Orsay, France}
\affiliation{$^{11}$ GSI Helmholtzzentrum f\"ur Schwerionenforschung, D-64291 Darmstadt, Germany}
\affiliation{$^{12}$ Department of Physics, P.O.~Box 3000, Fin-90014 University of Oulu, Finland}
\affiliation{$^{13}$ Institut Laue-Langevin, F-38042 Grenoble, France}
\affiliation{$^{14}$ Oliver Lodge Laboratory, University of Liverpool, Liverpool, L69 7ZE, United Kingdom}
\affiliation{$^{15}$ KTH-Royal Institute of Technology, SE-10044 Stockholm, Sweden}
\affiliation{$^{16}$ Institute of Physics, Slovak Academy of Sciences, Bratislava 845 11, Slovakia}
\affiliation{$^{17}$ Joint Institute of Nuclear Research, 141980 Dubna, Moscow Region, Russia}
\affiliation{$^{18}$ Centre d'Etudes Nucl{\'e}aires de Bordeaux Gradignan, F-33175 Gradignan, France}
\affiliation{$^{19}$ Service de Physique Nucl{\'e}aire Th{\'e}orique, Universit{\'e} Libre de Bruxelles, B-1050 Bruxelles, Belgium}

\date{\today}

\begin{abstract}
In-source resonant ionization laser spectroscopy of the even-$A$ polonium isotopes $^{192-210,216,218}$Po has been performed using the $6p^37s$ $^5S_2$ to $6p^37p$ $^5P_2$ ($\lambda=843.38$ nm) transition in the polonium atom (Po-I) at the CERN ISOLDE facility. The comparison of the measured isotope shifts in $^{200-210}$Po with a previous data set allows to test for the first time recent large-scale atomic calculations that are essential to extract the changes in the mean-square charge radius of the atomic nucleus. When going to lighter masses, a surprisingly large and early departure from sphericity is observed, which is only partly reproduced by Beyond Mean Field calculations.
\end{abstract}


\maketitle

In spite of substantial progress in understanding the structure of light atomic nuclei using \emph{ab-initio} calculations \cite{Mar09}, heavy nuclei are still modeled using shell-model calculations, symmetry-based approaches or mean-field descriptions. These concepts address an interesting aspect of the finite ensemble of strongly interacting fermions that makes the atomic nucleus, namely the subtle interplay between microscopic (individual nucleon) and macroscopic properties (collective behavior). The lead isotopes are a good example: while the doubly closed-shell nucleus $^{208}$Pb, with $Z=82$ and $N=126$, is a textbook case of a shell-model nucleus \cite{Sor08}, signatures of particle-hole symmetries have been found \cite{Cak05}. Furthermore, shape coexistence has been observed in the neutron-deficient isotopes, whereby states with different shapes occur at low excitation energy \cite{And00}. In spite of the latter effect, recent studies have shown that the ground state of the very neutron-deficient lead isotopes remains essentially spherical, confirming the robustness of the $Z=82$ shell closure in those isotopes \cite{DeW07,Sel09}. In the neighboring mercury isotopes ($Z=80$), a significant deviation from sphericity has been deduced around $N=104$, mid way between $N=82$ and $N=126$, from, e.g., the measurement of the changes in the mean-square charge radius ($\delta\langle r^2\rangle$) \cite{Ulm86}. From symmetry arguments a similar effect is expected to occur in the neutron-deficient polonium isotopes ($Z=84$) \cite{Hey83,Woo92}.

By approaching $N=104$, the intrusion of a presumed oblate band was seen at low energy in $^{192}$Po \cite{Hel96,Fot97}. In $^{190}$Po, evidence for a prolate configuration in the polonium isotopes was obtained \cite{Wis07}, which is believed to become the ground state in $^{188}$Po \cite{VdV03:PRC,And06}. In order to understand the change in the configuration, a direct measurement of ground-state properties independent of nuclear models is crucial. The $\delta\langle r^2\rangle$ can be extracted from isotope shifts measurements using laser spectroscopy in a model-independent procedure, provided that the atomic parameters (electronic factor $F$ and specific mass shift $M_{SMS}$) are known \cite{Fla10}. However, since there is no stable isotope of polonium, those measurements are extremely challenging. In particular, the information on the atomic structure of the polonium atom itself was hitherto rather scarce and limited in precision \cite{Cha66}. Hence a previous study of the $\delta\langle r^2\rangle$ in $^{200-210}$Po by laser spectroscopy \cite{Kow91} had to rely on a nuclear model (Finite Range Droplet Model - FRDM \cite{Mye83}) in order to extract the $\delta\langle r^2\rangle$ from the isotope shifts.

In the last 20 years, the progress in atomic structure theory has enabled large-scale calculations for open-shell atoms and ions \cite{Fri02} providing the necessary input to analyze laser-spectroscopic data. However, one still has to assess the accuracy and reliability of those calculations, especially in heavy systems, as they are needed to access the atomic and chemical properties of the super heavy elements.

In this Letter, we report on the measurement of the isotope shifts of the neutron-deficient, even-$A$ polonium isotopes from $^{210}$Po down to $^{192}$Po ($T_{1/2}=33$ ms) and of the neutron-rich isotopes $^{216,218}$Po using in-source resonant photoionization spectroscopy. Combining the isotope shifts of the present and previous data set \cite{Kow91} allowed us to test large-scale atomic  calculations for the first time and to evaluate their accuracy. The measured shifts and calculated atomic parameters are then used to extract $\delta\langle r^2\rangle$ and those are discussed in terms of nuclear models. 

The polonium isotopes were produced at the CERN ISOLDE facility by 1.4 GeV proton-induced spallation reactions of a thick depleted UC$_x$ target as described in Ref.~\cite{Coc10} over two experimental campaigns using the resonant ionization laser ion source \cite{Coc08}: $^{194-204}$Po in 2007 and $^{192,196,206-210,216,218}$Po in 2009. For $A>204$, the surface-ionized francium isotopes overwhelmed the beam. The $A>204$ polonium isotopes were instead studied without proton irradiation, using precursors previously accumulated in the target. Without proton irradiation, the short-lived, fast-releasing francium contamination disappears as it does not have any precursor. The isotopes $^{206-210}$Po were obtained via the $\beta^+$/EC decay of the isobaric astatine nuclei while the isotopes $^{216,218}$Po were produced via the $\alpha$ decay of $^{224}$Ra and $^{222}$Rn, respectively.

 \begin{figure}
  \begin{center}
   \includegraphics[width=\columnwidth]{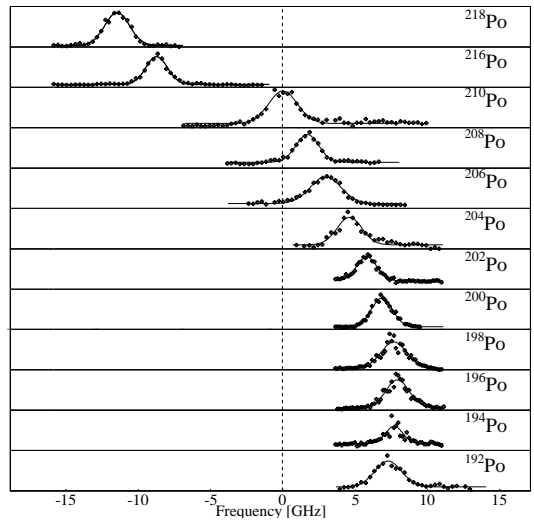}
   \caption{\label{fig:spectra}From top to bottom: example of a laser scan around the $843.38$ nm transition between the $6p^37s$ $^5S_2$ and $6p^37p$ $^5P_2$ atomic excited states in even-$A$ Po-I for $A=218$ down to $A=192$. For $A=192$, several scans have been added together to obtain sufficient statistics. The line is the result from a fit of those spectra using an asymmetric Voigt profile; more information on this asymmetry can be found in the text.
   }
  \end{center}
 \end{figure}

The $\alpha$-decaying isotopes $^{192-196,216,218}$Po were studied as described in Ref.~\cite{Coc10}. The $\beta$-decaying isotopes $^{200-204}$Po were sent to the ISOLDE tape station and their decay measured using a plastic scintillator and a single coaxial HPGe detector. The isotopes $^{206-210}$Po were directly counted in a Faraday cup. The measurements were performed by scanning the narrowband laser at $843.38$ nm from the $6p^37s$ $^5S_2$ to the $6p^37p$ $^5P_2$ excited state of the ionization scheme \cite{Coc08} and monitoring the yields as a function of the frequency. The frequency scans are shown in Fig.~\ref{fig:spectra}. The shape of the resonance can be described as a deformed Voigt profile. The 2007 data were analyzed as described in Ref.~\cite{Sel09}. For the 2009 data, an asymmetry in the fit function was introduced through a different Lorentzian width parameter on each side of the resonance. The difference comes from the use of different pump lasers between the two runs. The deduced isotope shifts with respect to $^{210}$Po ($\delta\nu^{A,210}_{exp}$) are presented in Table \ref{tbl:is}.

 \begin{table}
  \caption{\label{tbl:is}Isotope shifts $\delta\nu^{A,210}_{exp}$ in the $843.38$ nm transition of Po-I and changes in the mean-square charge radii $\delta\langle r^2\rangle^{A,210}_{exp}$ of the polonium isotopes with respect to $^{210}$Po from this work. The systematic uncertainty on the $\delta\langle r^2\rangle^{A,210}_{exp}$ originates from the Specific Mass Shift.}
  \begin{ruledtabular}
   \begin{tabular}{ccc}
    Mass & $\delta\nu^{A,210}_{exp}$ [GHz] & $\delta\langle r^2\rangle^{A,210}_{exp}$ [fm$^2$]\footnotemark[1] \\
    \hline
    ${218}$ & $-11.524(125)$ & $0.958(10)\{7\}$ \\
    ${216}$ & $-8.820(110)$ & $0.733(10)\{5\}$ \\
    ${210}$ & $0$ & $0$ \\
    ${208}$ & $1.631(120)$ & $-0.134(10)\{2\}$ \\
    ${206}$ & $3.042(120)$ & $-0.250(10)\{3\}$ \\
    ${204}$ & $4.419(110)$ & $-0.363(9)\{6\}$ \\
    ${202}$ & $5.724(140)$ & $-0.470(12)\{8\}$ \\
    ${200}$ & $6.828(110)$ & $-0.560(9)\{11\}$ \\
    ${198}$ & $7.570(165)$ & $-0.619(12)\{13\}$ \\
    ${196}$ & $7.733(100)$ & $-0.630(5)\{15\}$ \\
    ${194}$ & $7.361(155)$ & $-0.596(10)\{20\}$ \\
    ${192}$ & $6.697(185)$ & $-0.537(13)\{22\}$ \\
   \end{tabular}
  \end{ruledtabular}
  \footnotetext[1]{(statistical uncertainty)$\{$systematic uncertainty$\}$}
 \end{table}

\begin{figure}
  \begin{center}
   \includegraphics[width=\columnwidth]{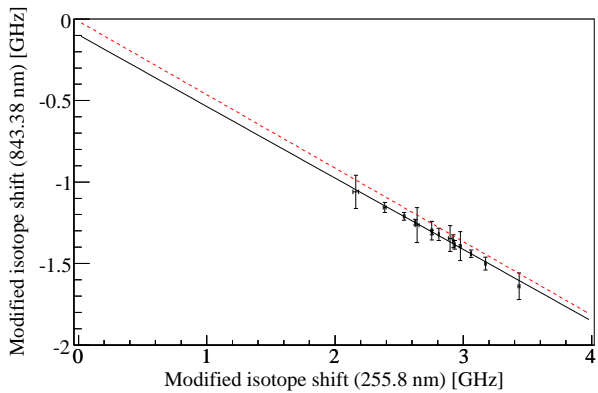}
   \caption{\label{fig:King}(Color online) King plot between the transitions at $255.8$ nm (\cite{Kow91}, $x$ axis) and at $843.83$ nm (present work, $y$ axis) for $^{200-210}$Po. The solid line is a linear fit through the data points; the dotted red line is the calculated relation from the large-scale atomic calculation and lies within $1$ $\sigma$ of the fit.}
  \end{center}
 \end{figure}

The large overlap in the studied isotopes between the present data set and the previous study using the $255.8$ nm transition \cite{Kow91} allows the formalism of King to be used to study the atomic properties of polonium by comparing modified isotope shifts (see Fig.~\ref{fig:King}) \cite{Fla10}. The slope of this so-called King plot is the ratio of the two $F$ while the $y$ intercept is a linear combination of the $M_{SMS}$ from each transition. The $M_{SMS}$ and $F$ can also be calculated on the basis of the Dirac-Coulomb-Breit Hamiltonian and a Fermi-like distribution of the different isotopes. A series of relativistic configuration interaction calculations have been carried out with systematically enlarged wave function expansions within a restricted active space, including the polarization of the electronic core and single and double excitations into three additional layers of correlation orbitals ($n=8,9,10$). Reasonable convergence with the size of the wave functions was obtained, especially for $F$, while the $M_{SMS}$ values appear more sensitive to the details of the calculations. The results are shown in Table \ref{tbl:atomic}. The red dotted line in Fig.~\ref{fig:King} displays the corresponding relation according to the calculated parameters and lies within $1$ $\sigma$ of the fitted trend. The good agreement between the slope of the calculated parameters and that of the fit shows the predictive power of the calculations for $F$. The difference in the $y$ intercept, however, raises some questions on the theoretical accuracy of the calculated $M_{SMS}$. 

 \begin{table}
  \caption{\label{tbl:atomic}Calculated atomic electronic factors $F$ and specific mass shifts $M_{SMS}$.}
  \begin{ruledtabular}
   \begin{tabular}{ccc}
    Transition [nm] & $F$ [GHz/fm$^2$] & $M_{SMS}$ [GHz$\cdot$amu] \\
    \hline
    $255.8$ & $28.363$ & $51$ \\
    $843.38$ & $-12.786$ & $-311$
   \end{tabular}
  \end{ruledtabular}
 \end{table}

\begin{figure}
  \begin{center}
   \includegraphics[width=\columnwidth]{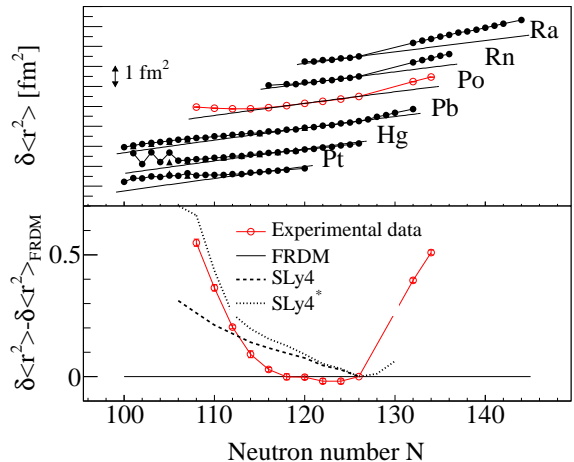}
   \caption{\label{fig:dr2}(Color online) (top) $\delta\langle r^2\rangle$ for the even-$Z$ isotopes from platinum ($Z=78$) to radium ($Z=88$) \cite{LeB99,Ulm86,Ans86,Din87,Dut91,DeW07,Fri05}. The solid black line represents the predictions from the spherical FRDM \cite{Mye83} using the second parametrization from Ref.~\cite{Ber85}. (bottom) Difference between the measured $\delta\langle r^2\rangle$ to the spherical FRDM. The dotted lines represent the Beyond Mean Field calculations with the SLy4 and SLy4$^{*}$ (with reduced pairing) interactions \cite{Ben04,Ben06}.}
  \end{center}
 \end{figure}

The $\delta\langle r^2\rangle$ of the even-$A$ isotopes $^{192-210,216,218}$Po were extracted using those parameters and a $0.932$ correction for higher moments \cite{Tor85}. In order to take into account the uncertainty of the different $M_{SMS}$ in the calculation, a systematic error was introduced. It was deduced as the difference between the $\delta\langle r^2\rangle$ values using only the calculated atomic parameters for the $843.38$ nm line and those obtained via the King plot and the calculated atomic parameters from the $255.8$ nm transition. The $\delta\langle r^2\rangle$ (see Table \ref{tbl:is}) are compared with the predictions from the spherical FRDM \cite{Mye83} using the second parametrization from Ref.~\cite{Ber85} (see Fig.~\ref{fig:dr2}). On the neutron-deficient side, a surprisingly large deviation from sphericity can be seen starting from $^{198}_{114}$Po that becomes increasingly marked for the lighter isotopes. The deviation is larger in magnitude and occurs for larger neutron numbers than in the $Z\leq82$ isotones. The data in the neutron-deficient radon and radium isotopes \cite{Fri05} do not extend far enough in the neutron-deficient side to compare with the polonium isotopes.

In order to understand the unexpectedly large and early deviation from sphericity in the polonium isotopes, the $\delta\langle r^2\rangle$ have been calculated using the same Beyond Mean Field method as in Ref.~\cite{Ben04,Ben06}. The most important feature of the method for this study is that the ground-state wave function is constructed as a superposition of mean-field wave functions corresponding to a large set of axial quadrupole deformations, projected on angular momentum and particle number. The coefficients of the expansion are determined by varying the energy corresponding to a Skyrme energy density functional. The SLy4 Skyrme parametrization has been tested together with the effect of a reduced pairing strength (SLy4$^{*}$). Within this framework, one cannot assign an intrinsic deformation to the wave functions. Instead, they are a mixture of mean-field states of different deformation and therefore different radii. In general, deformed configurations have larger radii than spherical ones. The two main effects that increase the radii of neutron-deficient polonium isotopes, compared to the global $A^{1/3}$ trend set by spherical configurations, are the spread of the collective wave function in deformation space and the shift of the dominant configurations from near-spherical to oblate. The increasing softness of the deformation energy surfaces, when going down from $^{210}$Po to $^{194}$Po, leads to collective ground-state wave functions of increasing spread, but which remain centered around spherical shapes. For $^{192,190}$Po, the ground-state wave function becomes centered around an oblate minimum in the deformation energy surface and the contribution from near-spherical configurations (or smaller radii) becomes suppressed.

The calculated $\delta\langle r^2\rangle$ are compared with the experimental data at the bottom of Fig.~\ref{fig:dr2}, after subtraction of the FRDM value. There is a qualitative agreement between theory and experiment, although deficiencies exist, especially for $^{192,194}$Po, where the data indicate a stronger deviation from sphericity. The effect of a reduced pairing strength, which clearly improves the agreement between theory and experiment for the lightest nuclei, is meant to put a larger weight on deformed oblate configurations. One still needs to construct more flexible energy functionals to correct the deficiencies of the actual ones.

Finally, the neutron-rich isotopes $^{216,218}$Po show a clear break from the trend of the polonium isotopes below $N=126$. The magnitude of this kink is similar to what is observed in the neutron-rich neighboring lead ($Z=82$) \cite{Ans86}, bismuth ($Z=83$) \cite{Pea00} and heavier isotopes. The underlying mechanism is still an open question.

In conclusion, in-source resonant ionization laser spectroscopy has been performed on the polonium isotopes from $^{192}$Po to $^{218}$Po. The overlap with the previous data set available in the literature has allowed testing of the large-scale atomic calculations for the $F$ and the $M_{SMS}$, although the latter remain slightly less well determined for such heavy atoms. This first experimental evidence of the reliability of such calculations is crucial for the study of laser spectroscopy in complex systems \cite{Fla10}. Moreover, it shows that reliable information can be extracted for the very heavy elements, where limited or no atomic data are available yet.

The $\delta\langle r^2\rangle$ of the even-$A$ polonium isotopes $^{192-210,216,218}$Po have been extracted and compared with systematics of this region and recent calculations. An unexpectedly large departure from sphericity was observed compared with the $Z\leq82$ isotones. Comparison to Beyond Mean Field calculations indicate that the coexistence of the different shapes at low excitation energies leads to a very soft nature of the most neutron-deficient polonium nuclei. The different trend with respect to the $Z\leq82$ nuclei might suggest that high-$j$ orbitals occupied by the protons play a critical role. The study of the more neutron-deficient radon ($Z=86$) and radium ($Z=88$) isotopes could eventually shed more light on the aspect.

\begin{acknowledgments}
 We would like to thank the ISOLDE collaboration for providing excellent beams and the GSI Target Group for manufacturing the carbon foils.
 This work was supported by FWO-Vlaanderen (Belgium), by GOA/2004/03 (BOF-K.U.Leuven), by the IUAP - Belgian State  Belgian Science Policy - (BriX network P6/23), by the European Commission within the Sixth Framework Programme through I3-EURONS (Contract RII3-CT-2004-506065), by the U.K.~Science and Technology Facilities Council, by the FiDiPro programme of the Finnish Academy and by the Slovak grant agency VEGA (Contract No.~1/0091/10).
\end{acknowledgments}

\end{document}